\documentclass[12pt]{article}
\usepackage{enumerate}
\usepackage{amsfonts}
\usepackage{amsmath}
\usepackage{amssymb}
\usepackage{amsthm}

\def\H{\mathcal{H}}

\newcommand{\Tr}{\mathrm{Tr}}

\newcounter{defin}  \newcounter{lemma}  \newcounter{theorem}
\newcounter{property} \newcounter{corol}  \newcounter{remark} \newcounter{example}
\newenvironment{lemma}{\par\refstepcounter{lemma}     \textbf{Lemma \thelemma.} }{\rm\par}
\newenvironment{theorem}{\par\refstepcounter{theorem}     \textbf{Theorem \thetheorem.}\ }{\rm\par}
\newenvironment{property}{\par\refstepcounter{property}     \textbf{Proposition \theproperty.}\ }{\rm\par}
\newenvironment{corollary}{\par\refstepcounter{corol}     \textbf{Corollary \thecorol.} }{\rm\par}

\newenvironment{remark}{\par\refstepcounter{remark}     \textbf{Remark \theremark.}}{\rm\par}

\begin{document}

\title{On the entanglement-assisted classical capacity of
infinite-dimensional quantum channels\thanks{
Work partially supported by RFBR grant 12-01-00319 and by the RAS research
program. The first author acknowledges support of the Cariplo Fellowship under the auspices of the Landau Network - Centro Volta.}}
\author{A.S. Holevo, M.E. Shirokov \\
Steklov Mathematical Institute, RAS, Moscow\\
holevo@mi.ras.ru, msh@mi.ras.ru}
\date{}
\maketitle

\begin{abstract}
The coding theorem for the entanglement-assisted
communication via infinite-dimensional quantum channel with linear
constraint is extended to a natural degree of generality. Relations between the entanglement-assisted classical
capacity and the $\chi$-capacity of constrained channels are obtained and
conditions for their coincidence are given. Sufficient conditions for
continuity of the entanglement-assisted classical capacity as a function of
a channel are obtained. Some applications of the obtained results to
analysis of Gaussian channels are considered.

A general (continuous) version of the fundamental relation between the
coherent information and the measure of privacy of classical information
transmission by infinite-dimensional quantum channel is proved.
\end{abstract}

\section{Introduction}

A central role in quantum information theory is played by the notion of
quantum channel -- the noncommutative analog of transition probability
matrix in classical theory. Informational properties of a quantum channel
are characterized by a number of different capacities depending on the type
of transmitted information, by additional resources used to increase the
rate of transmission, security requirements etc. see e.g. \cite{H-SSQT}. One
of the most important of these quantities is the entanglement-assisted
classical capacity, which characterizes the ultimate rate of classical
information transmission assuming that transmitter and receiver may use
common entangled state. By the definition this capacity is greater than or
equal to the classical (unassisted) capacity of the channel. The
Bennett-Shor-Smolin-Thaplyal (BSST) theorem~\cite{BSST} gives an explicit
expression for the entanglement-assisted capacity of a finite-dimensional
unconstrained channel showing that this capacity is equal to the maximum of
quantum mutual information.

When applying the protocol of entanglement-assisted communication to \emph{
infinite-dimensional} channels one has to impose certain constraints on the
input states. A typical physically motivated constraint is the bounded
energy of states used for encoding. This constraint is determined by the
linear inequality
\begin{equation}
\mathrm{Tr}\rho F\leq E,\quad E>0,  \label{lc}
\end{equation}
where $F$ is a positive self-adjoint operator -- a Hamiltonian of the input
quantum system. Operational definition of the entanglement-assisted
classical capacity of an infinite-dimensional quantum channel with linear
constraint (\ref{lc}) is given in~\cite{H-c-w-c}, where the generalization of the BSST
theorem is proved under special restrictions on the channel and on the
constraint operator. Recent advances in the study of entropic
characteristics of infinite-dimensional quantum channels (in particular,
generalization of the notion of quantum conditional entropy \cite{Kuz}) make
it possible to establish the general version of the BSST theorem for a
channel with linear constraints without any simplifying assumptions. This
result constitutes the first part of the paper.

The second part is devoted to study of relations between
entanglement-assisted and unassisted classical capacities of infinite-dimensional constrained channel and conditions for their (non-)coincidence.
It is shown that under certain circumstances the coincidence of the above
capacities implies that the channel is essentially classical-quantum (see
e.g. \cite{H-SSQT} for the definition).

We also consider the problem of continuity of the entanglement-assisted
classical capacity as a function of channel. This question has a physical
motivation in the fact that preparing a quantum channel in a real experiment
is subject to unavoidable imprecisions. In the finite-dimensional case the
continuity of the entanglement-assisted classical capacity was proved in
\cite{D&S}. In infinite dimensions this capacity is not continuous in
general (it is only lower semicontinuous), however we suggest several
sufficient conditions for its continuity and consider some applications.

In the last section we prove infinite-dimensional generalization of the
identity, due to Schumacher and Westmoreland \cite{Sch}, which underlies the
fundamental connection between the quantum capacity and the privacy of
classical information transmission through a quantum channel.

In the Appendix we give auxiliary facts concerning Bosonic Gaussian channels.

\section{Preliminaries}

Let $\mathcal{H}$ be a separable Hilbert space, $\mathfrak{B}(\mathcal{H})$
the algebra of all bounded operators in $\mathcal{H}$ and $\mathfrak{B}_{+}(
\mathcal{H})$ the positive cone in $\mathfrak{B}(\mathcal{H}).$ Let $
\mathfrak{T}(\mathcal{H})$ be the Banach space of all trace-class operators
in $\mathcal{H}$ and $\mathfrak{S}(\mathcal{H})$ the closed convex subset of
$\mathfrak{T}(\mathcal{H})$ consisting of positive operators with unit trace
called \emph{states} \cite{H-SSQT,N&Ch}. We denote by $I_{\mathcal{H}}$ the
unit operator in a Hilbert space $\mathcal{H}$ and by $\mathrm{Id}_{\mathcal{
\mathcal{H}}}$ the identity transformation of the Banach space $\mathfrak{T}
( \mathcal{H })$.

A linear completely positive trace-preserving map $\Phi :\mathfrak{T}(
\mathcal{H}_{A})\rightarrow \mathfrak{T}(\mathcal{H}_{B})$ is called \emph{\
channel} \cite{H-SSQT,N&Ch}. By the Stinespring dilation theorem complete
positivity of $\Phi $ implies existence of a Hilbert space $\mathcal{H}_{E}$
and of an isometry $V:\mathcal{H}_{A}\rightarrow \mathcal{H}_{B}\otimes
\mathcal{H}_{E}$ such that\footnote{
Here and in what follows we write $\mathrm{Tr}_{\mathcal{H}_{X}}(\cdot )=
\mathrm{Tr}_{X}(\cdot )$ for brevity.}
\begin{equation}
\Phi[\rho]=\mathrm{Tr}_{E}V\rho V^{\ast },\quad \rho \in \mathfrak{\ T }(
\mathcal{H}_{A}).  \label{Stinespring-rep}
\end{equation}
The channel $\widehat{\Phi}:\mathfrak{T}(\mathcal{H}_{A})\rightarrow
\mathfrak{T} (\mathcal{H}_{E})$ defined as
\begin{equation}
\widehat{\Phi }[\rho ]=\mathrm{Tr}_{B}V\rho V^{\ast },\quad \rho \in
\mathfrak{T}(\mathcal{H}_{A}),  \label{c-channel}
\end{equation}
is called \emph{complementary} to the channel $\Phi $ \cite{H-c-c}. The
complementary channel is defined uniquely in the following sense: if $
\widehat{\Phi }^{\prime }:\mathfrak{T}(\mathcal{H}_{A})\rightarrow \mathfrak{
\ \ T}(\mathcal{H}_{E^{\prime }})$ is a channel defined by (\ref{c-channel})
via another isometry $V^{\prime }:\mathcal{H}_{A}\rightarrow \mathcal{H}
_{B}\otimes \mathcal{H}_{E^{\prime }}$ then there is a partial isometry $W:
\mathcal{H}_{E}\rightarrow \mathcal{H}_{E^{\prime }}$ such that
\begin{equation*}
\widehat{\Phi }^{\prime }[\rho ]=W\widehat{\Phi }[\rho]W^{\ast },\quad
\widehat{\Phi }[\rho ]=W^{\ast }\widehat{\Phi }^{\prime }[\rho]W,\quad \rho
\in \mathfrak{T}(\mathcal{H}_{A}).
\end{equation*}

Let $H(\rho )$ be the von Neumann entropy of the state $\rho $ and $H(\rho
\Vert \sigma )$ the quantum relative entropy of the states $\rho $ and $
\sigma $ \cite{L,N&Ch,O&P}.\medskip\ A finite collection of states $\{\rho
_{i}\}$ with the corresponding probability distribution $\{\pi _{i}\}$ is
called \emph{ensemble} and denoted $\{\pi _{i},\rho _{i}\}$. The state $\bar{
\rho}=\sum_{i}\pi _{i}\rho _{i}$ is called the \emph{average state} of the
ensemble $\{\pi _{i},\rho _{i}\}$.\medskip\ The $\chi $-quantity of an
ensemble $\{\pi _{i},\rho _{i}\}$ is defined as
\begin{equation*}
\chi (\{\pi _{i},\rho _{i}\})\doteq \sum_{i}\pi _{i}H(\rho _{i}\Vert \bar{
\rho})=H(\bar{\rho})-\sum_{i}\pi _{i}H(\rho _{i}),  %\label{chi-q}
\end{equation*}
where the second expression is valid under the condition $H(\bar{\rho}
)<+\infty $. We will also use the notation $\chi _{\Phi }(\{\pi _{i},\rho
_{i}\})=\chi (\{\pi _{i},\Phi[\rho _{i}]\}).$ The $\chi$-quantity can be
considered as a quantum analog of Shannon information which appears in the
expression for the classical capacity of a quantum channel (see below).\medskip

Let $F$ be a positive self-adjoint operator in $\mathcal{H}_{A}$. For any
state $\rho \in \mathfrak{S}(\mathcal{H}_{A})$ the value $\mathrm{Tr}\rho F$
(finite or infinite) is defined as $\sup_{n}\mathrm{Tr}\rho P_{n}FP_{n}$,
where $P_{n}$ is the spectral projector of $F$ corresponding to the interval
$[0,n]$.\medskip

We impose the linear constraint onto the input states $\rho ^{(n)}$ of the
channel $\Phi ^{\otimes n}$
\begin{equation}
\mathrm{Tr}\rho ^{(n)}F^{(n)}\leq nE,  \label{constraint}
\end{equation}
where
\begin{equation}
F^{(n)}=F\otimes \dots \otimes I+\dots +I\otimes \dots \otimes F.
\label{f-n}
\end{equation}
\medskip

An operational definition of the classical capacity of a quantum channel
with the linear constraint can be found in \cite{H-c-w-c}. We will need an
analytical expression, for which we first introduce the $\chi $-capacity of
the channel $\Phi $ with the constraint (\ref{constraint}):
\begin{equation*}
C_{\chi }(\Phi ,F,E)=\sup_{\rho:\mathrm{Tr}\rho F \leq E}C_{\chi }(\Phi ,\rho ),
%\label{chi-cap}
\end{equation*}
where
\begin{equation}
C_{\chi }(\Phi ,\rho )=\sup_{\sum_{i}\pi _{i}\rho _{i}=\rho }\chi _{\Phi
}(\{\pi _{i},\rho _{i}\})  \label{chi-fun}
\end{equation}
is the constrained $\chi $-capacity of the channel $\Phi $ at the state $
\rho $ (the supremum is over all ensembles with the average state $\rho $).
If $H(\Phi \lbrack \rho ])<+\infty $ then
\begin{equation}
C_{\chi }(\Phi ,\rho )=H(\Phi \lbrack \rho ])-\hat{H}_{\Phi }(\rho ),
\label{chi-fun+}
\end{equation}
where $\hat{H}_{\Phi }(\rho )=\inf_{\sum_{i}\pi _{i}\rho _{i}=\rho
}\sum_{i}\pi _{i}H(\Phi \lbrack \rho _{i}])$ is the $\sigma$-convex
hull of the function $\rho \mapsto H(\Phi \lbrack \rho ])$. Due to
concavity of this function the infimum can be taken over ensembles
of pure states. By the Holevo-Schumacher-Westmoreland (HSW) theorem
adapted to constrained channels (\cite[Proposition 3]{H-c-w-c}), the
classical capacity of the channel $\Phi $ with constraint
(\ref{constraint}) is given by the following regularized expression
\begin{equation*}
C(\Phi ,F,E)=\lim_{n\rightarrow +\infty }n^{-1}C_{\chi }(\Phi ^{\otimes
n},F^{(n)},nE),
\end{equation*}
where $F^{(n)}$ is defined in (\ref{f-n}).\medskip

Another important analog of the Shannon information which appears in
connection with the entanglement-assisted classical capacity (see the next
section) is the \emph{quantum mutual information}. In finite dimensions it
is defined for arbitrary state $\rho \in \mathfrak{S}(\mathcal{H}_{A})$ by
the expression (cf.\cite{A&C})
\begin{equation}
I(\rho ,\Phi )=H(\rho )+H(\Phi \lbrack \rho ])-H((\Phi \otimes \mathrm{Id}
_{R})[\hat{\rho}]),  \label{mi-}
\end{equation}
where $\mathcal{H}_{R}$ is a Hilbert space isomorphic to $\mathcal{H}_{A}$, $
\,\hat{\rho}\,$ is a purification of the state $\rho $ in the space $
\mathcal{H}_{A}\otimes \mathcal{H}_{R}$ so that $\rho =\mathrm{Tr}_{R}\hat{
\rho}$. By using the complementary channel, the quantum mutual information
can be also expressed as follows
\begin{equation}
I(\rho ,\Phi )=H(\rho )+H(\Phi \lbrack \rho ])-H(\widehat{\Phi }[\rho ]).
\label{mi}
\end{equation}
In infinite dimensions expressions (\ref{mi-}),(\ref{mi}) may contain
uncertainty $"\infty -\infty "$, and to avoid this problem they should be
modified as
\begin{equation}
I(\rho ,\Phi )=H\left( (\Phi \otimes \mathrm{Id}_{R})[\hat{\rho}]\Vert (\Phi
\otimes \mathrm{Id}_{R})[\rho \otimes \varrho \,]\right) ,  \label{mi+}
\end{equation}
where $\varrho =\mathrm{Tr}_{A}\hat{\rho}$ is the state in $\mathfrak{S}(
\mathcal{H}_{R})$ with the same nonzero spectrum as $\rho $. Analytical
properties of the function $(\rho ,\Phi )\mapsto I(\rho ,\Phi )$ defined by (\ref{mi+})
were studied in \cite{H-Sh-4} in the infinite-dimensional case.

\section{The entanglement-assisted classical capacity}

Consider the following protocol of the classical information transmission
through the quantum channel $\Phi :\mathfrak{S}(\mathcal{H}_{A})\rightarrow
\mathfrak{S}(\mathcal{H}_{A^{\prime }})$.\footnote{
In this section the output of a quantum channel will be denoted $A^{\prime }$
for convenience of notations.} Two parties $A$ and $B$ share an entangled
(pure) state $\omega _{AB}$. $A$ does an encoding $\lambda \rightarrow
\mathcal{E}_{\lambda }$ of the classical signal $\lambda $ from the finite
alphabet $\Lambda $ with probabilities $\pi _{\lambda }$ and sends its part
of this shared state through the channel $\Phi $ to $B$. Here $\mathcal{E}
_{\lambda }$ are encoding channels depending on the signal $\lambda .$ Thus $
B$ gets the states $(\Phi \otimes \mathrm{Id}_{B})\left[ \omega _{\lambda }
\right] ,$ where $\omega _{\lambda }=(\mathcal{E}_{\lambda }\otimes \mathrm{
Id}_{B})\left[ \omega _{AB}\right] ,$ with probabilities $\pi _{\lambda },$
and $B$ aims to extract the maximum information about $\lambda $ by doing
measurements on these states. To enable block encodings, this picture should
be applied to the channel $\Phi ^{\otimes n}.$ Then the signal states $
\omega _{\lambda }^{(n)}$ transmitted through the channel $\Phi ^{\otimes
n}\otimes \mathrm{Id}_{B}^{\otimes n}$ have the special form
\begin{equation}
\omega _{\lambda }^{(n)}=(\mathcal{E}_{\lambda }^{(n)}\otimes \mathrm{Id}
_{B}^{\otimes n})\left[ \omega _{AB}^{(n)}\right] ,  \label{ABw}
\end{equation}
where $\omega _{AB}^{(n)}$ is the pure entangled state for $n$ copies of the
system $AB$ and $\lambda \rightarrow \mathcal{E}_{\lambda }^{(n)}$ are the
encodings for $n$ copies of the system $A$.\medskip

The constraint (\ref{constraint}) is equivalent to similar constraint onto
the input states of the channel $\Phi ^{\otimes n}\otimes \mathrm{Id}
_{B}^{\otimes n}$ with the constraint operator $F_{AB}^{(n)}=F^{(n)}\otimes
I_{B}^{\otimes n}$. Denote by $\mathcal{P}_{AB}^{(n)}$ the collection of
ensembles $\pi ^{(n)}=\{\pi _{\lambda }^{(n)},\omega _{\lambda }^{(n)}\},$
where $\omega _{\lambda }^{(n)}$ are states of the form (\ref{ABw}),
satisfying
\begin{equation*}
\sum_{\lambda \in \Lambda }\pi _{\lambda }^{(n)}\mathrm{Tr}\omega _{\lambda
}^{(n)}F_{AB}^{(n)}\leq nE.  %\label{constraint2}
\end{equation*}
The classical capacity of above protocol is called \emph{\
entanglement-assisted classical capacity }of the channel $\,\Phi $ under the
constraint (\ref{constraint}) and is denoted $C_{ea}(\Phi ,F,E)$ (for more
detail of the operational definition see \cite{H-c-w-c}). By a modification
of the proof of Proposition 2 in \cite{H-c-w-c},
\begin{equation}
C_{ea}(\Phi ,F,E)=\lim_{n\rightarrow \infty }\frac{1}{n}C_{ea}^{(n)}(\Phi
,F,E),  \label{cea}
\end{equation}
where
\begin{equation}
C_{ea}^{(n)}(\Phi ,F,E)=\sup_{\pi ^{(n)}\in \mathcal{\ }\mathcal{P}
_{AB}^{(n)}}\chi _{\Phi ^{\otimes n}\otimes \mathrm{Id}_{B}^{\otimes
n}}\left( \{\pi _{\lambda }^{(n)},\omega _{\lambda }^{(n)}\}\right) .
\label{cean}
\end{equation}
These are the expressions with which we will work in this paper. The
following Theorem generalizes Proposition 4 from \cite{H-c-w-c} to the case
of arbitrary channel $\Phi $ and arbitrary constraint operator $F$
.\smallskip

\begin{theorem}
\label{eac} \emph{Let $\,\Phi :\mathfrak{S}(\mathcal{H}_{A})\rightarrow
\mathfrak{S}(\mathcal{H}_{A^{\prime }})$ be a quantum channel and $F$ a
self-adjoint positive operator in the space $\mathcal{H}_{A}$. The
entanglement-assisted capacity (finite or infinite) of the channel $\,\Phi $
with the constraint (\ref{constraint}) is given by the expression}
\begin{equation}
C_{\mathrm{ea}}(\Phi ,F,E)=\sup_{\rho:\mathrm{Tr}\rho F \leq E}I(\rho ,\Phi ).
\label{eaco}
\end{equation}
\end{theorem}

It follows from Theorem \ref{eac} that the entanglement-assisted capacity
(finite or infinite) of the unconstrained channel $\,\Phi $ is
\begin{equation*}
C_{\mathrm{ea}}(\Phi )=\sup_{\rho \in \mathfrak{S}(\mathcal{H})}I(\rho ,\Phi
).
\end{equation*}

\textbf{Proof.} To prove the inequality $"\geq "$ in (\ref{eaco}) assume
first that the channel $\Phi $ has a finite-dimensional output (the system $
A^{\prime }$ is finite-dimensional). In this case the required inequality
can be proved by repeating the arguments from the corresponding part of the
proof of Proposition 4 in \cite{H-c-w-c} based on the special encoding
protocol. We only make the following remarks concerning generalization of
that proof:

\begin{itemize}
\item finite-dimensionality of the system $A^{\prime }$ implies finiteness
of the output entropy of the channel $\Phi$ on the whole space of input
states;

\item finiteness of the value $\mathrm{Tr} \rho F$ implies that all the
eigenvectors of the state $\rho$ belong to the domain of the operator $\sqrt{
F}$;

\item finite-dimensionality of the system $A^{\prime }$ shows that for any
finite rank state $\rho$ the restriction of the channel $\Phi^{\otimes n}$
to the support of the state $\rho^{\otimes n}$ acts as a finite-dimensional
channel for each $n$.

\item if there are no states satisfying the inequality $\mathrm{Tr}\rho F<E$
but there exists infinite rank state $\rho_0$ such that $\mathrm{Tr} \rho_0
F=E$ then there is a sequence $\{\rho_n\}$ of finite rank states converging
to $\rho_0$ such that $\mathrm{Tr} \rho_n F=E$ for which
\begin{equation*}
\liminf_{n\rightarrow+\infty} I(\rho_n, \Phi)\geq I(\rho_0, \Phi)
\end{equation*}
by lower semicontinuity of the quantum mutual information.
\end{itemize}

Let $\Phi $ be an arbitrary channel and $\{P_{n}\}$ be a sequence of finite-dimensional projectors in $\mathcal{H}_{A^{\prime }}$ strongly converging to
the unit operator $I_{A^{\prime }}$. The channel $\Phi $ is approximated in
the strong convergence topology (see \cite{H-Sh-3}) by the sequence of
channels $\Pi _{n}\circ \Phi $ with a finite-dimensional output, where $\Pi
_{n}(\rho )=P_{n}\rho P_{n}+[\mathrm{Tr}\rho (I_{A'}-P_{n})]\tau $ and $\tau $ is a
given state in $A^{\prime }$. Since the inequality $"\geq "$ in (\ref{eaco})
is proved for a channel with a finite-dimensional output, the chain rule for
the entanglement-assisted capacity implies
\begin{equation*}
C_{\mathrm{ea}}(\Phi ,F, E)\geq C_{\mathrm{ea}}(\Pi _{n}\circ \Phi , F,
E)\geq I(\rho ,\Pi _{n}\circ \Phi )\quad \text{for all}\; \rho \;\,\text{s.t.
}\;\, \mathrm{Tr}\rho F \leq E.
\end{equation*}
The lower semicontinuity of the function $\Phi \mapsto I(\rho ,\Phi )$ in
the strong convergence topology and the chain rule for the quantum mutual
information (see Proposition 1 in \cite{H-Sh-4}) imply
\begin{equation*}
\lim_{n\rightarrow +\infty }I(\rho ,\Pi _{n}\circ \Phi )=I(\rho ,\Phi )\leq
+\infty \quad \text{for all}\; \rho .
\end{equation*}
Hence the inequality $"\geq" $ in (\ref{eaco}) follows from the above
inequality.\medskip

We now prove the inequality $"\leq "$ in (\ref{eaco}). By Lemma \ref{bl}
below the expression $\chi _{\Phi ^{\otimes n}\otimes \mathrm{Id}
_{B}^{\otimes n}}(...)$ in the right hand side of (\ref{cean}) is bounded
from above by the quantity $I(\sum_{\lambda }\pi _{\lambda }^{(n)}(\omega
_{\lambda }^{(n)})_{A},\Phi ^{\otimes n})$. From~(\ref{cea}) we get
\begin{equation*}
C_{ea}(\Phi ,F,E)\leq \lim_{n\rightarrow \infty }\frac{1}{n}\sup_{\pi^{(n)}\in \mathcal{P}_{AB}^{(n)}}I\left( \sum_{\lambda }\pi
_{\lambda }^{(n)}(\omega _{\lambda }^{(n)})_{A},\Phi ^{\otimes n}\right) .
\end{equation*}
The right hand side is less than or equal to
\begin{equation*}
\sup_{\rho ^{(n)}:\mathrm{Tr}\rho ^{(n)}F^{(n)}\leq nE}I\left( \rho
^{(n)},\Phi ^{\otimes n}\right) \equiv \bar{I}_{n}(\Phi ).
\end{equation*}
But the sequence $\bar{I}_{n}(\Phi )$ is additive. To show this it is
sufficient to prove
\begin{equation}
\bar{I}_{n}(\Phi )\leq n\bar{I}_{1}(\Phi ).  \label{subad}
\end{equation}
By subadditivity of the quantum mutual information
\begin{equation*}
I\left( \rho^{(n)},\Phi ^{\otimes n}\right) \leq \sum_{j=1}^{n}I\left(
\rho_{j}^{(n)},\Phi \right) ,
\end{equation*}
where $\rho_{j}^{(n)}$ are the partial states, and by concavity,
\begin{equation*}
\sum_{j=1}^{n}I\left(\rho_{j}^{(n)},\Phi \right) \leq nI\left(
\frac{1}{n}\sum_{j=1}^{n}\rho_{j}^{(n)},\Phi \right) .
\end{equation*}
But $\mathrm{Tr}\rho^{(n)}F^{(n)}\leq nE$ is equivalent to $\mathrm{Tr}
\left( \frac{1}{n}\sum_{j=1}^{n}\rho_{j}^{(n)}\right) F\leq E,$ hence (\ref{subad}) follows. Thus
\begin{equation*}
C_{ea}(\Phi ,F,E)\leq \sup_{\rho :\mathrm{Tr}\rho F\leq E}I\left( \rho ,\Phi
\right) .
\end{equation*}

\begin{lemma}
\label{bl} \textit{Let $\,\Phi :\mathfrak{S}(\mathcal{H}_{A})\rightarrow
\mathfrak{S}(\mathcal{H}_{A^{\prime }})$ be a quantum channel and $\sigma $
an arbitrary state in $\,\mathfrak{S}(\mathcal{H}_{B})$. Then for an
arbitrary ensemble $\{\pi _{i},\omega _{i}\}$ of states in $\mathfrak{S}(
\mathcal{H}_{A}\otimes \mathcal{H} _{B})$ such that $\,(\omega
_{i})_{B}=\sigma \in \mathfrak{S}( \mathcal{H}_{B})\,$ for all $\,i$, the
following inequality holds
\begin{equation}
\chi _{\Phi \otimes \mathrm{Id}_{B}}\left( \{\pi _{i},\omega_{i}\}\right)
\leq I(\omega _{A},\Phi ),  \label{basic-ineq}
\end{equation}
where $\,\omega =\sum_{i}\pi _{i}\omega _{i}\,$ is the average state of the
ensemble $\,\{\pi _{i},\omega _{i}\}$.}
\end{lemma}

\vspace{5pt}

In the proof of this lemma we will use the infinite-dimensional
generalization of the conditional entropy proposed in \cite{Kuz} and briefly
described below.

In finite dimensions the conditional entropy of a state $\rho$ of a
composite system $AB$ is defined as
\begin{equation}  \label{c-e}
H(A|B)_{\rho}\doteq H(\rho)-H(\rho_B).
\end{equation}
The conditional entropy is finite, but in contrast to the classical case it
may be negative.

According to \cite{Kuz} the conditional entropy of a state $\rho$ of an
infinite-dimensional composite system $AB$ is defined as follows
\begin{equation}  \label{g-c-e}
H(A|B)_{\rho}\doteq H(\rho_A)-H(\rho\|\rho_A\otimes\rho_B)
\end{equation}
provided $H(\rho_A)<+\infty$. It is easy to see that the right hand sides of
(\ref{c-e}) and (\ref{g-c-e}) coincide if $H(\rho)<+\infty$ (finiteness of
any two values from the triple $H(\rho_A), H(\rho_B), H(\rho)$ implies
finiteness of the third one).

It is proved in \cite{Kuz} that the above-defined conditional entropy is a
concave function on the convex set of all states $\rho $ of the system $AB$
such that $H(\rho _{A})<+\infty $, possessing the following properties:
\begin{equation}
H(A|B)_{\rho _{AB}}\geq H(A|BC)_{\rho }\;  \label{mon}
\end{equation}
for any state$\;\rho \;$of $ABC$ (monotonicity), and
\begin{equation}
H(A|B)_{\rho _{AB}}=-H(A|C)_{\rho _{AC}}  \label{pure}
\end{equation}
$\;$for any pure state$\;\rho \;$of $ABC,$ where it is assumed that $
\;H(\rho _{A})<+\infty $. \medskip

\textbf{Proof of the lemma.} Let $\,\{\pi_i,\omega_i\}$ be an ensemble of
states in $\mathfrak{S}( \mathcal{H}_A\otimes\mathcal{H}_B)$ with the
average state $\omega$ such that $\,(\omega_i)_B=\sigma\in \mathfrak{S}(
\mathcal{H}_B)\,$ for all $\,i$. We have to show that
\begin{equation}  \label{basic-ineq+}
\sum_i \pi_i H\left(\Phi\otimes \mathrm{Id}_{B}[\omega_i]\,\|\,\Phi\otimes
\mathrm{Id}_{B}[\omega]\right)\leq I(\omega_A,\Phi).
\end{equation}

Let us prove first the inequality (\ref{basic-ineq+}) assuming that $\dim
\mathcal{H}_{A^{\prime }}<+\infty $ and $\dim \mathcal{H}_{B}<+\infty $. In
this case the left hand side of this inequality can rewritten as follows
\begin{equation*}
L\doteq H(\Phi \otimes \mathrm{Id}_{B}[\omega])-\sum_{i}\pi _{i}H(\Phi
\otimes \mathrm{Id}_{B}[\omega _{i}]).
\end{equation*}
By subadditivity of the von Neumann entropy we have
\begin{equation*}
L\leq H(\Phi[\rho])+\sum_{i}\pi _{i}[H(\sigma )-H(\Phi \otimes \mathrm{Id}
_{B}[\omega _{i}])],
\end{equation*}
where $\rho =\omega _{A}$. Note that $\,H(\Phi \otimes \mathrm{Id}
_{B}[\omega _{i}])-H(\sigma )\,$ is the conditional entropy $H(A^{\prime
}|B) $ at the state $\Phi \otimes \mathrm{Id}_{B}(\omega _{i})$. Let $\hat{
\omega}_{i}$ be a pure state in $ABR_{i}$ such that $(\hat{\omega}
_{i})_{AB}=\omega _{i}$. By monotonicity of the conditional entropy
(property (\ref{mon})) we have
\begin{equation}
H(\Phi \otimes \mathrm{Id}_{B}[\omega _{i}])-H(\sigma )=H(A^{\prime
}|B)_{\Phi \otimes \mathrm{Id}_{B}[\omega _{i}]}\geq H(A^{\prime
}|BR_{i})_{\Phi \otimes \mathrm{Id}_{BR_{i}}[\hat{\omega}_{i}]},  \label{one}
\end{equation}
where $H(A^{\prime }|BR_{i})$ is defined by (\ref{g-c-e}) (the system $R_{i}$
is infinite-dimensional, but the system $A^{\prime }$ is finite-dimensional
by the assumption). Since $\hat{\omega}_{i}$ is a purification of the state $
\rho _{i}\doteq (\omega _{i})_{A}$, i.e. $(\hat{\omega}_{i})_{A}=\rho _{i}$,
property (\ref{pure}) of the conditional entropy implies
\begin{equation}
\begin{array}{c}
H(A^{\prime }|BR_{i})_{\Phi \otimes \mathrm{Id}_{BR_{i}}[\hat{\omega}
_{i}]}=H(A^{\prime }|BR_{i})_{\mathrm{Tr}_{E}V\otimes I_{BR_{i}}\,\cdot \,
\hat{\omega}_{i}\,\cdot \,V^{\ast }\otimes I_{BR_{i}}} \\
\\
=-H(A^{\prime }|E)_{\mathrm{Tr}_{BR_{i}}V\otimes I_{BR_{i}}\,\cdot \,\hat{
\omega}_{i}\,\cdot \,V^{\ast }\otimes I_{BR_{i}}}=-H(A^{\prime }|E)_{V\rho
_{i}V^{\ast }},
\end{array}
\label{two}
\end{equation}
where $E$ is an environment system for the channel $\Phi $ and $V$ is the
Stinespring isometry (i.e. $\,\Phi[\rho]=\mathrm{Tr}_{E}V\rho V^{\ast }$).

By using concavity of the conditional entropy (defined by (\ref{g-c-e})) and
the property (\ref{pure}), we obtain
\begin{equation*}
\sum_i \pi_i H(A^{\prime}|E)_{V\rho_i V^*}\leq H(A^{\prime}|E)_{V\rho
V^*}=-H(A^{\prime}|R)_{\mathrm{Tr}_{E}V\otimes I_R\,\cdot\,\hat{\rho}
\,\cdot\, V^*\otimes I_R},
\end{equation*}
where $R$ is a reference system for the state $\rho$ and $\hat{\rho}$ is a
pure state in $AR$ such that $\hat{\rho}_A=\rho$. Hence (\ref{one}) and (\ref
{two}) imply
\begin{equation*}
L\leq H(\Phi[\rho])-H(A^{\prime}|R)_{\Phi\otimes\mathrm{Id}_R[\hat{\rho}
]}=H(\Phi\otimes\mathrm{Id}_R[\hat{\rho}]\|\Phi[\rho]\otimes\hat{\rho}
_R)=I(\rho, \Phi),
\end{equation*}
where definitions (\ref{mi+}) and (\ref{g-c-e}) were used.

Thus inequality (\ref{basic-ineq+}) is proved under the assumption $\dim
\mathcal{H}_{A^{\prime}}<+\infty$, $\dim\mathcal{H}_B<+\infty$. Its proof in
the general case can be obtained by means of an approximation technique as
follows.

Let $\{\pi_i,\omega_i\}$ be an ensemble such that $\,(\omega_i)_B= \sigma\in
\mathfrak{S}(\mathcal{H}_B)\,$ and $Q_n$ be the spectral projector of the
state $\sigma$ corresponding to its $n$ maximal eigenvalues. Let $\lambda_n=
\mathrm{Tr} Q_n\sigma$ and $C_n=I_A\otimes Q_n$. For a natural $n$ consider
the ensemble $\{\pi_i,\omega^n_i\}$ with the average state $\omega^n $,
where
\begin{equation*}
\begin{array}{c}
\omega^n_i=\lambda_n^{-1}C_n\omega_iC_n,\quad\omega^n=\lambda_n^{-1}C_n
\omega C_n.
\end{array}
\end{equation*}

Let $\{P_{n}\}$ be a sequence of finite-rank projectors in the space $
\mathcal{H}_{A^{\prime }}$, strongly converging to the identity operator $
I_{A^{\prime }}$, and $\tau $ a pure state in $\mathfrak{S}(\mathcal{H}
_{A^{\prime }})$. Consider the sequence of channels $\Phi _{n}=\Pi _{n}\circ
\Phi $, where
\begin{equation*}
\Pi _{n}[\rho]=P_{n}\rho P_{n}+\tau \mathrm{Tr}(I_{A^{\prime }}-P_{n})\rho
,\quad \rho \in \mathfrak{S}(\mathcal{H}_{A^{\prime }}).
\end{equation*}
Since $(\omega _{i}^{n})_{B}=\lambda _{n}^{-1}Q_{n}\sigma $ for all $i$, the
first part of the proof implies
\begin{equation*}
\sum_{i}\pi _{i}H\left( \Phi _{n}\otimes \mathrm{Id}_{B}[\omega
_{i}^{n}]\,\Vert \,\Phi _{n}\otimes \mathrm{Id}_{B}[\omega ^{n}]\right) \leq
I(\omega _{A}^{n},\Phi _{n}).
\end{equation*}
Since $\lambda _{n}\omega _{A}^{n}\leq \omega _{A}$, Lemma 4 in \cite{H-Sh-4}
shows that $\lim_{n\rightarrow +\infty }I(\omega _{A}^{n},\Phi
_{n})=I(\omega _{A},\Phi )$. Hence the above inequality implies inequality (\ref{basic-ineq+})
by lower semicontinuity of the relative entropy. This
proves the Lemma and completes the proof of the Theorem. $\square $

\section{Relations between entanglement-assisted and unassisted classical
capacities}

When dealing with infinite-dimensional quantum systems and channels, it is
necessary to consider \emph{generalized ensembles} defined as Borel
probability measures $\mu $ on the set of all quantum states. From this
point of view ordinary ensembles are described by finitely supported
measures $\mu .$ We denote by $\mathcal{P}(\mathfrak{S}(\mathcal{H}))$ the
set of all generalized ensembles of states in $\mathfrak{S}(\mathcal{H})$.

The $\chi $-quantity of a generalized ensemble $\mu $ is defined as
\begin{equation}
\chi (\mu )=\int_{\mathfrak{S}(\mathcal{H})}H(\rho \Vert \bar{\rho}(\mu
))\mu (d\rho )=H(\bar{\rho}(\mu ))-\int_{\mathfrak{S}(\mathcal{H})}H(\rho
)\mu (d\rho ),  \label{chi-q-g}
\end{equation}
where $\bar{\rho}(\mu )=\int_{\mathfrak{S}(\mathcal{H})}\rho \mu (d\rho )$
is the average state of $\mu $ (the Bochner integral) and the second formula
is valid under the condition $H(\bar{\rho}(\mu ))<+\infty $ \cite{H-Sh-2}.
For an arbitrary generalized ensemble $\mu\in\mathcal{P}(\mathfrak{S}(
\mathcal{H}_A))$ and a channel $\Phi:\mathfrak{S}(\mathcal{H}_A)\rightarrow
\mathfrak{S}(\mathcal{H}_B)$ one can define the new ensemble $\mu \circ \Phi
^{-1}\in\mathcal{P}(\mathfrak{S}(\mathcal{H}_B))$ (the image of the ensemble $\mu$ under action of the channel $\Phi$) as follows
\begin{equation*}
\mu \circ \Phi ^{-1}(B)=\mu (\left\{\rho\in \mathfrak{S}(\mathcal{H}_A)\,|\,
\Phi [\rho]\in B\right\} ).
\end{equation*}
The $\chi $-quantity of the ensemble $\mu \circ \Phi ^{-1}$ will be denoted $
\chi _{\Phi }(\mu )$. We have
\begin{equation}
\begin{array}{c}
\displaystyle\chi _{\Phi }(\mu )=\int_{\mathfrak{S}(\mathcal{H}_{A})}H(\Phi
[\rho]\Vert \Phi [\bar{\rho}(\mu )])\mu (d\rho ) \\
\displaystyle=H(\Phi [\bar{\rho}(\mu )])-\int_{\mathfrak{S}(\mathcal{H}
_{A})}H(\Phi [\rho])\mu (d\rho ),
\end{array}
\label{chi-phi-mu}
\end{equation}
where the second formula is valid under the condition $H(\Phi[\bar{\rho}
(\mu )])<+\infty $.\medskip

It is shown in \cite{H-Sh-2} that the constrained $\chi$-capacity defined by
(\ref{chi-fun}) can be expressed as follows
\begin{equation}
C_{\chi }(\Phi ,\rho )=\sup_{\mu:\bar{\rho}(\mu )=\rho }\chi _{\Phi }(\mu )
\label{chi-fun-g}
\end{equation}
(the supremum is over all generalized ensembles in $\mathcal{P}(\mathfrak{S}(
\mathcal{H}_A))$ with the average state $\rho $) and hence
\begin{equation}
C_{\chi }(\Phi ,F,E)=\sup_{\mu:\mathrm{Tr}\bar{\rho}(\mu )F\leq E}\chi _{\Phi
}(\mu ).  \label{chi-cap+}
\end{equation}

In this section we study the general relations between the capacities $
C_{\chi }(\Phi ,F,E)$, $C(\Phi ,F,E)$, $C_{\mathrm{ea}}(\Phi ,F,E)$ and give
conditions for their coincidence under the assumption\footnote{One can show this assumption holds if and only if $\mathrm{\Tr}\exp(-\lambda F)<+\infty$ for some $\lambda>0$.}
\begin{equation}
H(\rho )<+\infty \;\;\text{for all}\;\,\rho \;\,\text{such that}\;\,\mathrm{
\ \ Tr}\rho F \leq E,  \label{b-a}
\end{equation}
which implies, in particular, finiteness of all these values. A basic role
in this analysis is played by the following expression for the quantum
mutual information
\begin{equation}
I(\rho, \Phi)=H(\rho )+C_{\chi }(\Phi ,\rho )-C_{\chi }(\widehat{\Phi }
,\rho ),  \label{mi++}
\end{equation}
valid under the condition $H(\rho )<+\infty $ (since $C_{\chi }(\Phi ,\rho
)\leq H(\rho )$ for any channel $\Phi ,$ this condition implies finiteness
of all terms in the right-hand side of (\ref{mi++})).

If $H(\Phi [\rho])$ and $H(\widehat{\Phi }[\rho])$ are finite, then the
expression (\ref{mi++}) follows directly from (\ref{chi-fun+}) and (\ref{mi}),
since $\hat{H}_{\Phi }\equiv \hat{H}_{\widehat{\Phi }}$ (this follows
from the coincidence of $H(\Phi [\rho])$ and $H(\widehat{\Phi }[\rho])$ for
pure states $\rho $); in the general case it can be proved by using
Proposition \ref{inf-dim-l} in Section 6. \medskip

By subadditivity of the quantum mutual information expression (\ref{mi++})
implies the formal proof of the inequality
\begin{equation}  \label{one-two+}
C(\Phi ,F,E)\leq C_{\mathrm{ea}}(\Phi ,F,E),
\end{equation}
which looks obvious from the operational definitions of the capacities. It
also implies the following inequalities. \medskip

\begin{property}
\label{simple} \emph{Let $\,\Phi :\mathfrak{S}(\mathcal{H}_{A})\rightarrow
\mathfrak{S}(\mathcal{H}_{B})$ be a quantum channel and $F$ a positive
operator such that condition (\ref{b-a}) is valid. The following
inequalities hold}
\begin{equation}  \label{one+two}
\begin{array}{ccc}
C_{\mathrm{ea}}(\Phi ,F,E) & \geq & 2C_{\chi }(\Phi ,F,E)-C_{\chi }(\widehat{
\Phi },F,E), \\
&  &  \\
C_{\mathrm{ea}}(\Phi ,F,E) & \geq & 2C(\Phi ,F,E)-C(\widehat{\Phi },F,E),
\end{array}
\end{equation}
\emph{where $\widehat{\Phi }:\mathfrak{S}(\mathcal{H}_{A})\rightarrow
\mathfrak{S}(\mathcal{H}_{E})$ is the complementary channel to the channel $
\,\Phi $.}
\end{property}

\medskip

Note that in contrast to (\ref{one-two+}), both inequalities in (\ref{one+two}) hold with an equality if $
\Phi $ is the noiseless channel. These inequalities show that coincidence of
$C_{\mathrm{ea}}(\Phi ,F,E)$ with $C_{\chi }(\Phi ,F,E)$ (with $C(\Phi ,F,E)$) can take place only if
$C_{\chi }(\Phi ,F,E)\leq C_{\chi }(\widehat{\Phi },F,E)$ (correspondingly $
C(\Phi ,F,E)\leq C(\widehat{\Phi },F,E)$).  \medskip

\textbf{Proof.} For an arbitrary $\varepsilon >0$ let $\rho _{\varepsilon }$
be a state in $\mathfrak{S}(\mathcal{H}_{A})$ such that
\begin{equation*}
C_{\chi }(\Phi ,F,E)<C_{\chi }(\Phi ,\rho _{\varepsilon })+\varepsilon
,\quad \mathrm{Tr}\rho_{\varepsilon}F\leq E.
\end{equation*}
Since $C_{\chi }(\Phi ,\rho _{\varepsilon })\leq H(\rho _{\varepsilon
})<+\infty $, Theorem \ref{eac} and formula (\ref{mi++}) show that
\begin{eqnarray*}
 C_{\mathrm{ea}}(\Phi ,F,E) &\geq & I(\rho _{\varepsilon },\Phi) \\
&\geq & 2C_{\chi
}(\Phi ,\rho _{\varepsilon })-C_{\chi }(\widehat{\Phi },\rho _{\varepsilon }) \\
&\geq & 2C_{\chi }(\Phi ,F,E)-C_{\chi }(\widehat{\Phi },F,E)-2\varepsilon ,
\end{eqnarray*}
which implies the first inequality in (\ref{one+two}).

The second inequality in (\ref{one+two}) is obtained from the first one by
regularization. $\square$ \medskip

Now we consider the question of coincidence of the capacities $C_{\mathrm{ea}
}(\Phi ,F,E)$ and $C_{\chi }(\Phi ,F,E)$.\medskip

We call a channel $\Phi :\mathfrak{S}(\mathcal{H}_{A})\rightarrow \mathfrak{
\ S }(\mathcal{H}_{B})$ \emph{classical-quantum} (briefly, \emph{c-q channel}
) if the image of the dual channel $\Phi ^{\ast }:\mathfrak{T}(\mathcal{H}
_{B})\rightarrow \mathfrak{T}(\mathcal{H}_{A})$ consists of commuting
operators. If all these operators are diagonal in a fixed orthonormal basis $
\{|k\rangle \}$ in $\mathcal{H}_{A},$ we say that the c-q channel is of
\emph{discrete type}. In this case it has the following representation
\begin{equation}
\Phi [\rho]=\sum_{k=1}^{\dim \mathcal{H}_{A}}\langle k|\rho |k\rangle \sigma
_{k},  \label{c-q-rep}
\end{equation}
where $\{\sigma _{k}\}$ is a collection of states in $\mathfrak{S}(\mathcal{\H }_{B})$. Any finite dimensional c-q channel
is of the discrete type. An example of a c-q channel which is not of the discrete type
is provided by Bosonic Gaussian c-q channel (see Appendix).\medskip

It is shown in \cite{Sh-19} that $\,C_{\chi }(\Phi )=C_{\mathrm{ea}}(\Phi )$
for any finite-dimensional unconstrained c-q channel $\Phi $ and that this
equality \emph{implies} that the restriction of the channel $\Phi $ to the
support of the average state of any optimal ensemble is a c-q channel (an
ensemble is called optimal if its $\chi$-quantity coincides with $C_{\chi
}(\Phi )$, see \cite{Sch-West}). The example from \cite{E} shows that the
words "the restriction of" in the last assertion can not be removed.

To generalize the above assertion to infinite dimensions we have to consider
the notion of a generalized optimal ensemble for a constrained infinite-dimensional channel introduced in \cite{H-Sh-2}.

A generalized ensemble $\mu_* $ is called optimal for the channel $\Phi $
with constraint (\ref{constraint}) if
\begin{equation*}
\mathrm{Tr}\bar{\rho}(\mu_*)F\leq E\quad \mathrm{and}\quad \,C_{\chi }(\Phi
,F,E)=\chi _{\Phi }(\mu_* ),
\end{equation*}
which means that the supremum in (\ref{chi-cap+}) is achieved on $\mu_* $.

This is a natural generalization of the notion of the optimal ensemble for
finite-dimensional (constrained or unconstrained) channel. In contrast to
the finite-dimensional case, an optimal generalized ensemble for infinite-dimensional constrained channel may not exist, but one can prove the
following sufficient condition for existence of such ensemble.\medskip

\begin{property}
\label{opt-ens} \cite{H-Sh-2} \emph{If the subset of $\,\mathfrak{S}(\mathcal{\H}_A)$ defined by the inequality $\mathrm{Tr}\rho F\leq E$ is compact
\footnote{
This subset is compact if and only if the spectrum of operator $F$ consists of eigenvalues of finite multiplicity,
accumulating at infinity, see the Lemma in \cite{H-c-w-c} and Lemma 3
in \cite{H-Sh-2}.} and the function $\rho\mapsto H(\Phi[\rho])$ is
continuous on this subset then there exists a generalized optimal ensemble
for the channel $\,\Phi$ with constraint (\ref{constraint}).}
\end{property}\smallskip

This condition holds for arbitrary Bosonic Gaussian channel with energy
constraint \cite{H-Sh-2}, the remark after Proposition 3. It also holds for
any channel having the Kraus representation with a finite number of summands
provided the operator $F$ satisfies the condition $\mathrm{Tr}\exp (-\lambda
F)<+\infty $ for all $\lambda >0$ (this can be proved by using Proposition
6.6 in \cite{O&P}).

The following theorem gives a necessary condition for coincidence of the
capacities $C_{\chi }(\Phi ,F,E)$ and $C_{\mathrm{ea}}(\Phi ,F,E)$.\medskip

\begin{theorem}
\label{coincidence} \emph{Assume that there exist a generalized optimal
ensemble $\mu_*$ for the channel $\,\Phi:\mathfrak{S}(\mathcal{H}_A)\rightarrow \mathfrak{S}(\mathcal{H}_B)$
with constraint (\ref{constraint})
(in particular, the condition of Proposition \ref{opt-ens} holds) and that
condition (\ref{b-a}) is valid. Let $\,\mathcal{H}_{\ast}$ be the support of
the average state of $\mu_*$, i.e. $\,\mathcal{H}_{\ast}=\mathrm{supp}\bar{
\rho}(\mu_*)$.} \smallskip

\emph{If $\,C_{\chi }(\Phi ,F,E)=C_{\mathrm{ea}}(\Phi ,F,E)$ then the
restriction of the channel $\,\Phi $ to the set $\,\mathfrak{S}(\mathcal{H}
_{\ast})$ is a c-q channel of discrete type.} \smallskip
\end{theorem}

\medskip

\textbf{Proof.} Without loss of generality we may assume that the optimal
generalized ensemble $\mu_*$ is supported by pure states. This follows from
convexity of the function $\sigma\mapsto H(\Phi[\sigma]\|\Phi[\rho])$, since
for an arbitrary measure $\mu\in\mathcal{P}(\mathfrak{S}(\mathcal{H}_A))$
there exists a measure $\hat{\mu}\in \mathcal{P}(\mathfrak{S}(\mathcal{H}_A))
$ supported by pure states such that $\bar{\rho}(\hat{\mu})=\bar{\rho}(\mu)$
and $\int f(\sigma)\hat{\mu} (d\sigma)\geq\int f(\sigma)\mu(d\sigma)$ for
any convex lower semicontinuous nonnegative function $f$ on $\mathfrak{S}(
\mathcal{H}_A)$ (this measure $\hat{\mu}$ can be constructed by using the
arguments from the proof of the Theorem in \cite{H-Sh-2}).\smallskip

The equality $C_{\chi }(\Phi ,F,E)=C_{\mathrm{ea}}(\Phi ,F,E)$ implies $
C_{\chi }(\Phi ,\bar{\rho}(\mu_* ))=I(\Phi ,\bar{\rho}(\mu_* ))$. It follows
from condition (\ref{b-a}) and representation (\ref{mi++}) that this is
equivalent to the equality $H(\bar{\rho}(\mu_* ))=C_{\chi }(\widehat{\Phi },
\bar{\rho}(\mu_* ))<+\infty $. Since $C_{\chi }(\Phi ,\bar{\rho}(\mu_*
))=\chi _{\Phi }(\mu_* )$, the remark after Proposition \ref{inf-dim-l} in
Section 6 and condition (\ref{b-a}) imply $C_{\chi }(\widehat{\Phi },\bar{
\rho}(\mu_* ))=\chi _{\widehat{\Phi }}(\mu_* )$. Since $H(\bar{\rho}(\mu_*
))=\chi (\mu_* )$, the equality $H(\bar{\rho}(\mu_* ))=\chi _{\widehat{\Phi }
}(\mu_* )$ shows that the channel $\widehat{\Phi }$ preserves the $\chi $
-quantity of the ensemble $\mu_* $, i.e. $\chi _{\widehat{\Phi }}(\mu_*
)=\chi (\mu_* )$. By Theorem 5 in \cite{Sh-20} the restriction of the
channel $\widehat{\widehat{\Phi }}\cong\Phi $ to the set $\mathfrak{S}(
\mathcal{H}_{\ast})$ is a c-q channel of discrete type. $\square $

\smallskip

\begin{remark}
\label{coincidence-r} In contrast to unconstrained channels the assertion of
Theorem \ref{coincidence} is not reversible even in finite dimensions: the
entanglement-assisted classical capacity of a discrete type c-q channel with
linear constraint may be greater than its unassisted classical capacity \cite
[ Example 3]{Sh-19}. By repeating the arguments from the proof of Theorem 2
in \cite{Sh-19} and using condition (\ref{b-a}) one can show that \emph{$
\,C_{\chi }(\Phi ,F,E)=C_{\mathrm{ea}}(\Phi ,F,E)$ for any discrete type c-q
channel $\,\Phi $ with constraint (\ref{constraint}) provided the operator $F
$ is diagonal in the basis $\{|k\rangle \}$ from representation (\ref
{c-q-rep}) of the channel $\,\Phi $}.
\end{remark}

\medskip

For an arbitrary nontrivial subspace $\mathcal{H}_0$ of $\mathcal{H}_A$ the
restriction of the channel $\Phi:\mathfrak{S}(\mathcal{H}_A)\rightarrow
\mathfrak{S}(\mathcal{H}_B)$ to the subset $\mathfrak{S}(\mathcal{H}_0)$
will be called \emph{subchannel} of $\Phi$ corresponding to the subspace $
\mathcal{H}_0$.\medskip

Theorem \ref{coincidence} implies the following sufficient condition for
non-coincidence of $C_{\mathrm{ea}}(\Phi ,F,E)$ and $\,C_{\chi}(\Phi ,F,E)$. \medskip

\begin{corollary}
\label{coincidence-c} \emph{Let the assumptions of Theorem \ref{coincidence}
hold. Then $\,C_{\mathrm{ea}}(\Phi ,F,E)>C_{\chi }(\Phi,F,E)$ if one of the
following conditions is valid:}

\begin{enumerate}
\item \emph{the channel $\,\Phi$ is not a c-q channel of discrete type and the
optimal measure $\mu_*$ has a non-degenerate average state;}

\item \emph{the channel $\,\Phi$ has no c-q subchannels of discrete type.}
\end{enumerate}
\end{corollary}

\medskip

As mentioned before the assumptions of Theorem \ref{coincidence} hold for
arbitrary Gaussian channel $\Phi_{K,l,\alpha}$ if $F$ is the Hamiltonian of
the input system ($K,l,\alpha$ are the parameters of the channel, see the
Appendix). By Corollary \ref{coincidence-c} and Proposition \ref{gauss} in
the Appendix the strict inequality $C_{\mathrm{ea}}(\Phi_{K,l,\alpha},F,E)>
\,C_{\chi}(\Phi_{K,l,\alpha},F,E)$ holds if one of the following conditions
is valid:

\begin{enumerate}
\item $K\neq0$ and the optimal measure $\mu_*$ has a non-degenerate average
state;

\item the rank of $K$ coincides with the dimension $2k$ of the input
symplectic space ($k$ is the number of the input modes).
\end{enumerate}

Condition 1 holds if the conjecture of Gaussian optimizers (see \cite{H-SSQT}
, Ch.12) is valid for the channel $\Phi_{K,l,\alpha}$ .

\section{On continuity of the entanglement-assisted capacity}

Since a physical channel is always determined with some finite accuracy, it
is necessary to explore the question of continuity of its information
capacity with respect to small perturbations of a channel. It means,
mathematically, that we have to study continuity of the capacity as a
function of a channel assuming that the set of all channels is equipped with
some appropriate topology.

In this section we consider continuity properties of the
entanglement-assisted capacity with respect to the strong convergence
topology on the set of all channels \cite{H-Sh-3}. Strong convergence of a
sequence of the channels $\Phi_n:\mathfrak{S}( \mathcal{H}_A)\rightarrow
\mathfrak{S}(\mathcal{H}_B)$ to the channel $\Phi_0: \mathfrak{S}(\mathcal{H}
_A)\rightarrow\mathfrak{S}(\mathcal{H}_B)$ means that $\lim_{n\rightarrow+
\infty}\Phi_n[\rho]=\Phi_0[\rho]$ for any state $\rho\in\mathfrak{S}(
\mathcal{H}_A)$.

Theorem \ref{eac} and lower semicontinuity of the quantum mutual information
as a function of a channel in the strong convergence topology imply that $
\Phi \mapsto C_{\mathrm{ea}}(\Phi ,F,E)$ is a lower semicontinuous function
in this topology on the set of all quantum channels, i.e.
\begin{equation*}
\liminf_{n\rightarrow +\infty }C_{\mathrm{ea}}(\Phi _{n},F,E)\geq C_{\mathrm{
ea}}(\Phi_{0},F,E)\;\;(\leq +\infty )
\end{equation*}
for any sequence $\{\Phi _{n}\}$ of channels strongly converging to the
channel $\Phi _{0}$.\smallskip

The following proposition gives sufficient conditions for the continuity.
\smallskip

\begin{property}
\label{cont-cond} \textit{Let $\,F$ be a self-adjoint positive operator such that $\mathrm{\ Tr}\exp(-\lambda F)<+\infty$ for all $\lambda>0$ and $\,\{\Phi_n\}$ be a
sequence of channels strongly converging to a channel $\,\Phi_0$. The relation
\begin{equation}  \label{lim-exp}
\lim_{n\rightarrow+\infty}C_{\mathrm{ea}}(\Phi_n,F,E)=C_{\mathrm{ea}
}(\Phi_0,F,E)<+\infty
\end{equation}
holds if one of following conditions is valid:}

\begin{enumerate}
\item \emph{$\lim_{n\rightarrow+\infty}H(\Phi_n[\rho_n])=H(\Phi_0[\rho_0])$
for an arbitrary sequence $\{\rho_n\}$ converging to a state $\rho_0$ such
that $\mathrm{Tr} \rho_n F\leq E$ for all $n=0,1,2,...$;}

\item \emph{there exists a sequence $\,\{\widehat{\Phi}_n\}$ of channels
strongly converging to a channel $\,\widehat{\Phi}_0$ such that $(\Phi_n,
\widehat{\Phi}_n)$ is a complementary pair for each $n=0,1,2,...$}
\end{enumerate}
\end{property}

\vspace{5pt}

Condition 1 in Proposition \ref{cont-cond} holds for any converging
sequence of Gaussian channels provided that $F$ is oscillator Hamiltonian of
a Bosonic system.

Condition 2 in Proposition \ref{cont-cond} holds for the sequence of the
channels
\begin{equation*}
\Phi_{n}[\rho]=\sum_{i=1}^{+\infty}V^{n}_{i}\rho(V_{i}^{n})^{*},
\end{equation*}
where $\{V^{n}_{i}\}_{n}$ is a sequence of operators from $\H_A$ into $\H_B$
strongly converging to the operator $V^{0}_{i}$ for each $i$ such that $
\sum_{i=1}^{+\infty}(V_{i}^{n})^{*}V^{n}_{i}=I_A$ for all $n$. Indeed,
\begin{equation*}
\widehat{\Phi}_{n}[\rho]=\sum_{i,j=1}^{+\infty}[\mathrm{Tr}
V_{i}^{n}\rho(V_{j}^{n})^{*}]|i\rangle\langle j|,
\end{equation*}
where $\{|i\rangle\}_{i=1}^{+\infty}$ is an orthonormal basis in $\mathcal{H}
_E$, and it is easy to see that the sequences $\{\Phi_n\}$ and $\{\widehat{
\Phi}_n\}$ strongly converge to the channels $\Phi_0$ and $\widehat{\Phi}_0$
(defined by the same formulas with $n=0$).\medskip

\textbf{Proof.} Note first that the set $\mathcal{A}=\{\rho\in\mathfrak{S}(
\mathcal{H}_A)\,|\,\mathrm{Tr}\rho F\leq E\}$ is compact (by the Lemma in
\cite{H-c-w-c}) and the function $\rho\mapsto H(\rho)$ is continuous on this
set (by Proposition 6.6 in \cite{O&P}).

By Proposition 4 in \cite{H-Sh-4} for each $n$ the function $\rho\mapsto
I(\rho,\Phi_n)$ is continuous on the compact set $\mathcal{A}$ and hence
\begin{equation*}
C_{\mathrm{ea}}(\Phi_n,F,E)=\sup_{\rho\in\mathcal{A}}I(\rho,\Phi_n)=I(
\rho_n, \Phi_n)
\end{equation*}
for a particular state $\rho_n$ in $\mathcal{A}$.

Assume that there exists
\begin{equation}  \label{lim-exp+}
\lim_{n\rightarrow+\infty}C_{\mathrm{ea}}(\Phi_n,F,E)>C_{\mathrm{ea}
}(\Phi_0,F,E)
\end{equation}
By the remark before Proposition \ref{cont-cond}, to prove (\ref{lim-exp})
it suffices to show that (\ref{lim-exp+}) leads to a contradiction.

Since the set $\mathcal{A}$ is compact, we can assume (by passing to a
subsequence) that the sequence $\{\rho_n\}$ converges to a particular state $
\rho_0\in\mathcal{A}$. Hence to obtain a contradiction to (\ref{lim-exp+})
it suffices to prove that
\begin{equation}  \label{lim-exp++}
\lim_{n\rightarrow+\infty}I(\rho_n,\Phi_n)=I(\rho_0,\Phi_0).
\end{equation}

The above conditions $1$ and $2$ provide two different ways to prove (\ref
{lim-exp++}).

If condition $1$ holds then
\begin{equation*}
I(\rho_n,\Phi_n)=H(\rho_n)+H(\Phi_n[\rho_n])-H(\Phi_n\otimes\mathrm{Id}
_R[|\varphi_n\rangle \langle\varphi_n|]),
\end{equation*}
where $|\varphi_n\rangle$ is any purification for the state $\rho_n$, $
n=0,1,2,...$.

By lower semicontinuity of the function $(\Phi,\rho)\mapsto I(\rho,\Phi)$,
continuity of the entropy on the set $\mathcal{A}$ and condition $1$, to
prove (\ref{lim-exp++}) it suffices to show that
\begin{equation*}
\liminf_{n\rightarrow+\infty}H(\Phi_n\otimes\mathrm{Id}_R[|\varphi_n\rangle
\langle\varphi_n|])\geq H(\Phi_0\otimes\mathrm{Id}_R[|\varphi_0\rangle
\langle\varphi_0|]).
\end{equation*}
This relation follows from lower semicontinuity of the relative entropy,
since strong convergence of the sequence $\{\Phi_n\}$ to the channel $\Phi_0$
implies strong convergence of the sequence $\{\Phi_n\otimes\mathrm{Id}_R\}$
to the channel $\Phi_0\otimes\mathrm{Id}_R$ and we can choose such sequence $
\{|\varphi_n\rangle\}$ that converges to the vector $|\varphi_0\rangle$ \cite
[Lemma 2]{H-Sh-4}.\smallskip

If condition $2$ holds then (\ref{lim-exp++}) directly follows from
Proposition 5 in \cite{H-Sh-4}.

\section{Coherent information and a measure of private classical information}

Let $\,\Phi :\mathfrak{S}(\mathcal{H}_{A})\rightarrow \mathfrak{S}(\mathcal{\H }_{B})$
be a quantum channel and $\,\widehat{\Phi }:\mathfrak{S}(\mathcal{H } _{A})\rightarrow \mathfrak{S}(\mathcal{H}_{E})$ be its
complementary channel. In finite dimensions the \emph{coherent information}
of the channel $\Phi $ at any state $\rho $ is defined as a difference
between $H(\Phi [\rho])$ and $H(\widehat{\Phi }[\rho])$ \cite{N&Ch,Sch}. The
coherent information is still another quantum analog of the Shannon
information relevant to the quantum capacity of the channel \cite
{H-SSQT,N&Ch}. In infinite dimensions the values $H(\Phi[\rho])$ and $H(
\widehat{\Phi }[\rho])$ may be infinite even for a state $\rho $ with finite
entropy, therefore the coherent information may be defined via the quantum
mutual information as a function with values in $(-\infty,+\infty]$ as follows (cf.\cite{H-Sh-4})
\begin{equation*}
I_{c}(\rho, \Phi)=I(\rho, \Phi)-H(\rho ).
\end{equation*}

Let $\rho $ be a state in $\,\mathfrak{S}(\mathcal{H}_{A})$ with finite
entropy. By monotonicity of the $\chi $-quantity the values $\chi _{\Phi
}(\mu )$ and $\chi _{\widehat{\Phi }}(\mu )$ do not exceed $H(\rho )=\chi
(\mu )$ for any measure $\mu \in \mathcal{P}(\mathfrak{S}( \mathcal{H}_{A}))$
supported by pure states with the barycenter $\rho $. The following
proposition can be considered as a generalization of the basic
relation in \cite{Sch} which underlies the fundamental connection between
the quantum capacity and the private transmission of the classical
information through a quantum channel. A measure for the latter
is given by the difference $\chi _{\Phi }(\mu )-\chi _{ \widehat{\Phi }}(\mu
)$ between the $\chi$-quantities of the receiver and the environment (the
eavesdropper).\medskip

\begin{property}
\label{inf-dim-l} \emph{Let $\mu$ be a measure in $\mathcal{P}( \mathfrak{S}
( \mathcal{H}_A))$ supported by pure states with the barycenter $\rho$. Then
\begin{equation}  \label{chi-dif}
\chi_{\Phi}(\mu)-\chi_{\widehat{\Phi}}(\mu)=I(\rho,\Phi)-H(\rho)=I_c(\rho,
\Phi).
\end{equation}
}
\end{property}

This proposition shows, in particular, that the difference $\chi _{\Phi
}(\mu )-\chi _{\widehat{\Phi }}(\mu )$ does not depend on $\mu $. So, if the
supremum in expression (\ref{chi-fun-g}) for the value $C_{\chi }(\Phi ,\rho )$ is achieved
at some measure $\mu _{\ast }$ then the
supremum in the similar expression for the value $C_{\chi }(\widehat{\Phi }
,\rho )$ is achieved at this measure $\mu _{\ast }$ and vice versa.
\smallskip

\textbf{Proof.} If $H(\Phi[\rho])<+\infty$ then $H(\widehat{\Phi}
[\rho])<+\infty$ by the triangle inequality (see \cite{N&Ch}) and (\ref
{chi-dif}) can be derived from (\ref{mi}) by using the second formula in (\ref{chi-phi-mu})
and by noting that the functions $\rho\mapsto H(\Phi[\rho])
$ and $\rho\mapsto H(\widehat{\Phi}[\rho])$ coincide on the set of pure
states. In general case we have to use the approximation method to prove (\ref{chi-dif}).
To realize this method it is necessary to introduce some
additional notions. \smallskip

Let $\mathfrak{T}_{1}(\mathcal{H})=\{A\in\mathfrak{T}(\mathcal{H})\,|\,A\geq
0,\;\mathrm{Tr} A\leq 1\}$. We will use the following two extensions of the
von Neumann entropy to the set $\mathfrak{T}_{1}(\mathcal{H})$ (cf.\cite{L})
\begin{equation*}
S(A)=-\mathrm{Tr} A\log A,\quad H(A)=S(A)+\mathrm{Tr} A\log \mathrm{Tr}
A,\quad \forall A \in\mathfrak{T}_{1}(\mathcal{H}).
\end{equation*}
Nonnegativity, concavity and lower semicontinuity of the von Neumann entropy
imply similar properties of the functions $S$ and $H$ on the set $\mathfrak{T}_{1}(\mathcal{H})$.

The relative entropy for two operators $A,B\in\mathfrak{T}_{1}( \mathcal{H})$
is defined as follows (see details in \cite{L})
\begin{equation*}
H(A\,\|B)=\sum_{i}\langle i|\,(A\log A-A\log B+B-A)\,|i\rangle,
\end{equation*}
where $\{|i\rangle\}$ is the orthonormal basis of eigenvectors of $A$. By
means of this extension of the relative entropy the $\chi$-quantity of a
measure $\mu$ in $\mathcal{P}(\mathfrak{T}_{1}(\mathcal{H}))$ is defined by
the first expression in (\ref{chi-q-g}).\footnote{$\mathcal{P}(\mathfrak{T}_{1}(\mathcal{H
}))$ is the set of all probability measures on $\mathfrak{T} _{1}(\mathcal{H}
)$ equipped with the weak convergence topology.}

A completely positive linear map $\Phi:\mathfrak{T}( \mathcal{H}
_A)\rightarrow\mathfrak{T}(\mathcal{H}_B)$, which does not increase trace,
is called \emph{\ quantum operation} \cite{N&Ch}. For any quantum operation $
\Phi$ the Stinespring representation (\ref{Stinespring-rep}) holds, in which
$V$ is a contraction. The complementary operation $\widehat{\Phi}:\mathfrak{T
}( \mathcal{H}_A)\rightarrow\mathfrak{T}(\mathcal{H}_E)$ is defined via this
representation by relation (\ref{c-channel}).

By the obvious modification of the arguments used in the proof of
Proposition 1 in \cite{H-Sh-2} it is easy to show that the function $
\mu\mapsto\chi(\mu)$ is lower semicontinuous on the set $\mathcal{P} (
\mathfrak{T}_{1}(\mathcal{H}))$ and that for an arbitrary quantum operation $
\Phi$ and a measure $\mu\in\mathcal{P} ( \mathfrak{S}(\mathcal{H}_A))$ such
that $S(\Phi[\bar{\rho}(\mu)])<+\infty$ the $\chi$-quantity of the measure $
\mu\circ\Phi^{-1}\in\mathcal{P}( \mathfrak{T}_{1}(\mathcal{H}_B))$ can be
expressed as follows
\begin{equation}  \label{formula}
\chi_{\Phi}(\mu)= S(\Phi[\bar{\rho}(\mu)])-\int_{\mathfrak{S}(\mathcal{H}
_A)}S(\Phi[\rho])\mu(d\rho).
\end{equation}

We are now in a position to prove (\ref{chi-dif}) in general case. Note that
for a given measure $\mu\in\mathcal{P}(\mathfrak{S}(\mathcal{H} _A)) $ the
function $\Phi\mapsto\chi_{\Phi}(\mu)$ is lower semicontinuous on the set of
all quantum operations equipped with the strong convergence topology (in
which $\Phi_n\rightarrow\Phi$ means $\Phi_n[\rho]\rightarrow \Phi[\rho]$ for
all $\rho$ \cite{H-Sh-3}). This follows from the lower semicontinuity of the
functional $\mu\mapsto\chi(\mu)$ on the set $\mathcal{\ P }(\mathfrak{T}_1(
\mathcal{H}_B))$, since for an arbitrary sequence $\{\Phi_n\}$ of quantum
operations strongly converging to a quantum operation $\Phi$ the sequence $
\{\mu\circ\Phi_n^{-1}\}$ weakly converges to the measure $\mu \circ\Phi^{-1}$
(this can be verified directly by using the definition of weak convergence
and by noting that for sequences of quantum operations the strong
convergence is equivalent to the uniform convergence on compact subsets of $
\mathfrak{S} (\mathcal{H}_A)$, see the proof of Lemma 1 in \cite{H-Sh-3}).

Let $\{P_n\}$ be an increasing sequence of finite-rank projectors in $
\mathfrak{B} (\mathcal{H}_B)$ strongly converging to $I_B$. Consider the
sequence of quantum operations $\Phi_n=\Pi_n \circ\Phi$, where $
\Pi_n[\sigma]=P_n[\sigma]P_n$. Then
\begin{equation}  \label{c-oper}
\widehat{\Phi}_n[\rho]=\mathrm{Tr}_{\mathcal{H}_B}P_n\otimes I_{\mathcal{H}
_E}V\rho V^*,\quad \rho\in\mathfrak{S}(\mathcal{H}_A),
\end{equation}
where $V$ is the isometry from Stinespring representation (\ref
{Stinespring-rep}) for the channel $\Phi$.

The sequences $\{\Phi _{n}\}$ and $\{\widehat{\Phi }_{n}\}$ strongly
converge to the channels $\Phi $ and $\widehat{\Phi }$ correspondingly. Let $
\rho =\sum_{k}\lambda _{k}|k\rangle \langle k|$ and $|\varphi _{\rho
}\rangle =\sum_{k}\sqrt{\lambda _{k}}|k\rangle \otimes |k\rangle $. Since $
H(\rho)<+\infty $ and $S(\Phi _{n}[\rho])<+\infty $, the triangle inequality
implies $S(\widehat{\Phi }_{n}[\rho])<+\infty $. So, we have
\begin{equation}
\begin{array}{c}
\displaystyle I(\rho,\Phi _{n})\doteq H\left(\Phi _{n}\otimes \mathrm{Id}
_{R}[|\varphi _{\rho }\rangle \langle \varphi _{\rho }|]\,\Vert \,\Phi
_{n}[\rho]\otimes \varrho \right) \\
\\
\displaystyle=-S(\widehat{\Phi }_{n}[\rho])+S(\Phi _{n}[\rho])+a_{n}=-\chi _{
\widehat{\Phi }_{n}}(\mu )+\chi _{\Phi _{n}}(\mu )+a_{n},
\end{array}
\label{I-n}
\end{equation}
where
\begin{equation}  \label{a-exp}
a_{n}=-\sum_{k} \mathrm{Tr}(\Phi _{n}[|k\rangle \langle k|])\lambda _{k}\log
\lambda _{k}\;
\end{equation}
and the last equality is obtained by using (\ref{formula}) and coincidence
of the functions $\rho \mapsto S (\Phi [\rho])$ and $\rho \mapsto S (
\widehat{\Phi }[\rho])$ on the set of pure states.

Since the function $\Phi\mapsto I(\rho,\Phi)$ is lower semicontinuous (by
lower semicontinuity of the relative entropy) and $I(\rho,\Phi_n)\leq
I(\rho,\Phi)$ for all $n$ by monotonicity of the relative entropy under the
action of the quantum operation $\Pi_n\otimes\mathrm{Id}_{R}$, we
have
\begin{equation}  \label{lim-r-1}
\lim_{n\rightarrow+\infty}I(\rho,\Phi_n)=I(\rho,\Phi).
\end{equation}

We will also prove that
\begin{equation}  \label{lim-r-2}
\lim_{n\rightarrow+ \infty} \chi_{\Phi_n}(\mu)= \chi_{\Phi}(\mu)\quad \text{
and}\quad \lim_{n\rightarrow+ \infty}\chi_{\widehat{\Phi}_n}(\mu)= \chi_{
\widehat{\Phi}}(\mu).\;
\end{equation}
The first relation in (\ref{lim-r-2}) follows from the lower semicontinuity
of the function $\Phi\mapsto\chi_{\Phi}(\mu)$ established before and from
the inequality $\chi_{\Phi_n}(\mu)\leq \chi_{\Phi}(\mu)$ valid for all $n$
by monotonicity of the $\chi$-quantity under the action of the quantum
operation $\Pi_n$.

To prove the second relation in (\ref{lim-r-2}) note that (\ref{c-oper})
implies $\widehat{\Phi}_n[\rho]\leq\widehat{\Phi}[\rho]$ for any state $
\rho\in \mathfrak{S}(\mathcal{H}_A)$. Hence Lemma 2 in \cite{H-Sh-3} shows
that
\begin{equation}  \label{chi-ineq}
\chi_{\widehat{\Phi}_{n}}(\mu)\leq \chi_{\widehat{\Phi}}(\mu)+ f(\mathrm{Tr}
\widehat{\Phi}_{n}[\rho]),
\end{equation}
where $f(x)=-2x\log x-(1-x) \log(1-x)$, for any measure $\mu\in\mathcal{P}(
\mathfrak{S}(\mathcal{H}_A))$ with finite support and the barycenter $\rho$.
Let $\mu$ be an arbitrary measure in $\mathcal{P}(\mathfrak{S}(\mathcal{H}
_A))$ with the barycenter $\rho$ and $\{\mu_k\}$ the sequence of measures with finite support and the
same barycenter constructed in the proof of Lemma 1 in \cite{H-Sh-2},
which weakly converges to the measure $\mu$. Validity of inequality (\ref{chi-ineq})
 for the measure $\mu$ is derived from its validity for all the
measures $\mu_k$ by using lower semicontinuity of the function $\mu\mapsto
\chi_{\widehat{\Phi}_{n}}(\mu)$ and the inequality $\chi_{\widehat{\Phi}}(\mu_k)\leq \chi_{\widehat{\Phi}}(\mu)$
valid for all $k$ by the construction of the sequence $\{\mu_k\}$ and convexity of the
relative entropy.

Inequality (\ref{chi-ineq}) and the lower semicontinuity of the function $
\Phi\mapsto\chi_{\Phi} (\mu)$ imply the second relation in (\ref{lim-r-2}).

Since $\{a_n\}$ defined in (\ref{a-exp}) obviously tends to $H(\rho)$,
relations (\ref{I-n}), (\ref{lim-r-1}) and (\ref{lim-r-2}) imply (\ref
{chi-dif}). $\square$

\section{Appendix: Gaussian classical-quantum channels}

The main applications of infinite-dimensional quantum information theory are
related to Bosonic system, for detailed description of which we refer to
Ch.12 in \cite{H-SSQT}. Let $\mathcal{H}_{A}$ be the irreducible
representation space of the Canonical Commutation Relations (CCR)
\begin{equation}
W_A(z_{A})W_A(z_{A}^{\prime })=\exp \left(-\frac{i}{2}z_{A}^{\top }\Delta
_{A}z_{A}^{\prime }\right) W_A(z_{A}^{\prime }+z_{A})  \label{CCR}
\end{equation}
with a coordinate symplectic space $(Z_{A},\Delta _{A})$ and the Weyl system
$W_{A}(z_A)=\exp (iR_{A}\cdot z_{A});\,z_{A}\in Z_{A}$. Here $R_{A}$ is the
row-vector of the canonical variables in $\mathcal{H}_{A}$, and $\Delta _{A}$
is the nondegenerate skew-symmetric commutation matrix of the components of $
R_{A}$. Gaussian channel $\Phi :\mathfrak{T}(\mathcal{H}_{A})\rightarrow
\mathfrak{T}(\mathcal{H}_{B}),$ with a similar description for $\mathcal{H}
_{B}$,\ is defined via the action of its dual $\Phi ^{\ast }$ on the Weyl
operators:\bigskip
\begin{equation}
\Phi ^{\ast }[W_{B}(z_{B})]=W_A(Kz_{B})\exp \left[ il^{\top }z_{B}-\frac{1}{2}
z_{B}^{\top }\alpha z_{B}\right] ,  \label{gaus-ch}
\end{equation}
where $K$ is matrix of a linear operator $Z_{B}\rightarrow Z_{A}$, $l\in
Z_{B}$ and $\alpha $ is real symmetric matrix satisfying

\begin{equation*}
\alpha \geq \pm \frac{i}{2}\left( \Delta _{B}-K^{\top }\Delta _{A}K\right) .
%\label{nid}
\end{equation*}

\begin{property}
\label{gauss} \emph{Let $\Phi_{K,l,\alpha}$ be the Gaussian channel with
parameters $K,l,\alpha$.}\smallskip

1) \emph{The channel $\,\Phi_{K,l,\alpha}$ is c-q if and only if
\begin{equation*}
K^{\top }\Delta _{A}K=0.  %\label{cl-qu}
\end{equation*}
In this case it is of discrete type if and only if $K=0$ i.e. the channel is
completely depolarizing.}\smallskip

2) \emph{If $\,rank K=dim Z_A$ then the channel $\,\Phi_{K,l,\alpha}$ has no
c-q subchannels of discrete type. }
\end{property}

\medskip

\textbf{Proof.} 1) Since the family $\{W_B(z_B)\}_{z_B\in Z_B}$ generates $\mathfrak{B}(\mathcal{H}_B)$,
all the operators $\Phi_{K,l,\alpha}^*(A)$, $A\in\mathfrak{B}(\mathcal{H}_B)$, commute if and only if operators (\ref{gaus-ch}),
i.e. $W_A(Kz_{B}),$ commute for all $z_{B}$. By (\ref{CCR}),
\begin{equation*}
W_A(Kz_{B})W_A(Kz_{B}^{\prime })=\exp \left(-iz_{B}^{\top }K^{\top }\Delta
_{A}Kz_{B}^{\prime }\right) W_A(Kz_{B}^{\prime })W_A(Kz_{B}),
\end{equation*}
hence the first assertion follows. Assuming the discrete representation (\ref
{c-q-rep}), we have that the operators $W_A(Kz_{B})=\exp (iR_{A}\cdot Kz_{B})$
all have pure point spectrum, which is possible only if $Kz_{B}\equiv 0$
since the canonical observables $R_{A}$ are known to have Lebesgue spectrum.

2) Suppose there is a subspace $\mathcal{H}_0\subset\mathcal{H}_A$ such that
\begin{equation*}
\Phi_{K,l,\alpha}[\rho]=\sum_k\langle k|\rho|k\rangle\sigma_k
\end{equation*}
for all $\rho\in\mathfrak{S}(\mathcal{H}_0)$, where $\{|k\rangle\}$ is an
orthonormal basis in $\mathcal{H}_0$. Then
\begin{equation*}
\begin{array}{c}
\displaystyle PW_A(Kz_B)P\exp \left[ il^{\top }z_{B}-\frac{1}{2} z_{B}^{\top }\alpha z_{B}
\right]=P\Phi_{K,l,\alpha}^*[W_B(z_B)]P\\\\\displaystyle=\sum_k [\mathrm{Tr} W_B(z_B)\sigma_k]|k\rangle\langle k|,
\end{array}
\end{equation*}
where $P=\sum_k |k\rangle\langle k|$ is the projector onto $\mathcal{H}_0$.
It follows that $\langle k|W_A(Kz_B)|j\rangle=0$ for all $k\neq j$. But this can
not be valid, since $\{Kz_B\,|\,z_B\in Z_B\}=Z_A$ and hence the family $
\{W_A(Kz_B)\}_{z_B\in Z_B}$ of Weyl operators acts irreducibly on $\mathcal{H}_A$.
$\square$

\end{document}